\documentclass[11pt]{article}
\usepackage{epsfig}

\newcommand{\dfrac}[2]{\frac{\displaystyle{#1}}{\displaystyle{#2}}}

\newcommand{\sech}{ {\rm sech}}

\newcommand{\sn}{ {\rm sn}}
\newcommand{\cn}{ {\rm cn}}
\newcommand{\dn}{ {\rm dn}}
\newcommand{\ns}{ {\rm ns}}
\newcommand{\nc}{ {\rm nc}}
\newcommand{\nd}{ {\rm nd}}
\newcommand{\sd}{ {\rm sd}}
\newcommand{\ds}{ {\rm ds}}
\newcommand{\cs}{ {\rm cs}}
\newcommand{\cd}{ {\rm cd}}
\newcommand{\dc}{ {\rm dc}}

\begin{document}
\baselineskip=18pt

\parskip 1.8mm
\begin{center}
{\Large\bf  Comments on: ``Optical solitons in a parabolic law media
with fourth order dispersion"\,[\,Appl. Math. Comput.
\,208\,(2009)\,209-302\,]}\\[12pt]
{Gui-Qiong Xu\footnote{E-mail address: \ xugq@staff.shu.edu.cn\,(G.-Q. Xu)}}\\[8pt]
{\small Department of Information Management, Shanghai University,
Shanghai 200444, PR China}
\end{center}

\vspace{6mm}
\begin{center}
\begin{minipage}{5.2in}
\baselineskip=20pt { \small {\bf Abstract.}\,Recently, Biswas and
Milovic\,[Appl. Math. Comput.\,208\,(2009)\,209-302] have found
optical one-soliton solutions of a fourth order dispersive
cubic-quintic nonlinear Schr\"odinger equation. In this comment, we
first show there are mistakes in the paper and demonstrate that the
obtained solutions do not satisfy the considered equation. And then
we reconstruct a series of analytical exact solutions by means of a
direct ansatz method and F-expansion method. These solutions include
solitary wave solutions of the bell shape, solitary wave solutions
of the kink shape, and periodic wave solutions of Jacobian elliptic
function.}
\end{minipage}
\end{center}

\section*{1.\,\,Analysis of the solutions given in
Ref.\cite{biswas_amc_2009}} \mbox{}\hskip 0.6cm As is well known,
the investigation for solition solutions of nonlinear Schr\"odinger
equation is always an important and attractive topic. Very recently,
Biswas and Milovic\cite{biswas_amc_2009} considered the higher order
dispersive cubic-quintic nonlinear Schr\"odinger equation,
\begin{equation}\label{hnls_01}
  i\,q_{t}\,+\,a\,q_{xx}\,-\,b\,q_{xxxx}\,+\,
  c\,(|q|^2\,+d\,|q|^4\,)\,q\,=\,0,
\end{equation}
and obtained the optical soliton solution of Eq.(\ref{hnls_01}).
However, we find there are mistakes in the
paper\cite{biswas_amc_2009} and the obtained solution does not
satisfy Eq.(\ref{hnls_01}).

Biswas {et al.}\cite{biswas_amc_2009} first introduced the
transformation,
\begin{equation}\label{hnls_02}
  q(x,t)\,=\,P(x,t)\,{\rm e}^{i\,(-\kappa\,x+\omega\,t+\theta)},
\end{equation}
where $P(x,t)$ is a real function to be determined later, and
$\kappa, \omega$ are real constants. By using the transformation
(\ref{hnls_02}), Eq.(\ref{hnls_01}) is converted into a complex
differential equation of $P(x,t)$, in which the real and imaginary
parts read,
\begin{equation}\label{hnls_03}
 \dfrac{\partial{P}}{\partial t}\,-\,2\kappa(a+2b\kappa^2)\, \dfrac{\partial{P}}{\partial x}
  +4b\,\kappa\,\dfrac{\partial^3{P}}{\partial x^3}\,=\,0,
\end{equation}
\begin{equation}\label{hnls_04}
(\omega+a\,\kappa^2+b\kappa^4)\,P\,-\,cP^3\,-\,c\,d\,P^5\,-\,(a+6b\kappa^2)\,
\dfrac{\partial^2{P}}{\partial x^2}
  +b\,\dfrac{\partial^4{P}}{\partial x^4}\,=\,0.
\end{equation}

Then the solution of Eqs.(\ref{hnls_03})-(\ref{hnls_04}) was
supposed as
\begin{equation}\label{hnls_05}
  P\,=\,\dfrac{A}{(\lambda+\cosh\tau)^p},
  \,\,\,\tau\,=\,B(x-\nu\,t).
\end{equation}
Substituting Eq.(\ref{hnls_05}) into
Eqs.(\ref{hnls_03})-(\ref{hnls_04}), the authors obtained two
expressions with respect to $\cosh\tau$ and $\sinh\tau$. We have
noticed that there are many mistakes about the expressions (12)-(13)
given in Ref.\cite{biswas_amc_2009}. Equating the coefficients of
$1/(\lambda+\cosh\tau)^{p+j}(j=0,\cdots,4)$ of the obtained
expressions, the values of $A, B, \omega, \lambda$ and $\nu$ were
determined.

At last, the authors obtained the optical soliton solution of
Eq.(\ref{hnls_01}) as follows,
\begin{equation}\label{hnls_06}
   q(x,t)\,=\,\dfrac{A}{\lambda+\cosh(B(x-\nu\,t))}\,{\rm
   e}^{i\,(-\kappa\,x+\omega\,t+\theta)},
\end{equation}
where $A, B, \omega, \lambda$ and $\nu$ were given by Eqs.(16)-(21)
of Ref.\cite{biswas_amc_2009}.

However, the ``solution" (\ref{hnls_06}) does not satisfy
Eq.(\ref{hnls_01}). We can note this fact without substituting
(\ref{hnls_06}) into Eq.(\ref{hnls_01}). The solitons are the
results of a delicate balance between dispersion and nonlinearity,
thus it is impossible that the linear partial differential equation
(\ref{hnls_03}) admits the bell type solitary wave (\ref{hnls_05}).

To be on the save side we have substituted Eq.(\ref{hnls_05}) with
$p=1$ into Eq.(\ref{hnls_03}) and have obtained the following
expression,
$$
  E_1\,=\,\left[\dfrac{\nu+2a\kappa+4b\kappa^3-4b\kappa B^2}{(\lambda+\cosh\tau)^2}\,+\,
  \dfrac{24b\kappa B^2\cosh \tau}{(\lambda+\cosh\tau)^3}\,-\,
  \dfrac{24b\kappa B^2\sinh^2\tau)}{(\lambda+\cosh\tau)^4}\right]\,A\,B\sinh\tau.
$$
We can see that this expression is equal to zero only in two cases.
One is $A=0$ or $B=0$, and the other is $\kappa=\nu=0$. This means
that the ``solution" (\ref{hnls_06}) obtained by Biswas et al. in
\cite{biswas_amc_2009} is not correct.

\section*{2.\,\,New optical solitary wave solution of Eq.(\ref{hnls_01})}

\mbox{}\hskip 0.6cm In the following, we adopt the ansatz solution
of Li {et al.}\cite{lizh_prl_2000} in the form
\begin{equation}\label{hnls_li01}
  q(x,t)\,=\,E(x,t)\,{\rm e}^{i\,(k\,x-\omega\,t+\theta)},
\end{equation}
where $E(x,t)$ is the complex envelope function, and $k, \omega$ are
real constants. Substituting Eq.(\ref{hnls_li01}) into
Eq.(\ref{hnls_01}) and removing the exponential term, we can rewrite
Eq.(\ref{hnls_01}) as
\begin{equation}\label{hnls_li02}
\begin{array}{l}
 i\,E_t\,+\,2ik\,(a+2b\,k^2)\,E_x\,+\,(a+6\,b\,k^2)\,E_{xx}\,-\,4i\,b\,k\,E_{xxx}\,-\,
 b E_{xxxx}\\[0.35cm]
 \,+\,(\omega-a\,k^2-b\,k^4)\,E\,+c\,|E|^2E\,+\,c\,d\,|E|^4E\,=\,0.
\end{array}
\end{equation}

We now take the complex envelope ansatz function $E(x,t)$ as
\begin{equation}\label{hnls_li03}
  E(x,t)\,=\,i\,\beta\,+\,\lambda\,\tanh(\xi),\,\,\xi=p\,x+s\,t,
\end{equation}
where $\beta, \lambda, p, s$ are real constants. Substituting
Eq.(\ref{hnls_li03}) into Eq.(\ref{hnls_li02}) and setting the
coefficients of $\tanh(\xi)^{j}(j=0,\cdots, 5)$ to zero, one obtains
the following algebraic equations:
\[
\begin{array}{l}

\lambda\,(dc{\lambda}^{4}-24\,{p}^{4}b )=0,\hskip 0.4cm

\lambda ({\lambda}^{3}\beta\,cd+24\,b{p}^{3}k
)=0,\\[0.3cm]

\lambda
(s-2\,\lambda\,{\beta}^{3}cd-\lambda\,\beta\,c+2\,kpa+32\,b{p}^
{3}k+4\,bp{k}^{3})=0,\\[0.3cm]

\lambda\, ({\lambda}^{2}c+2\,{\lambda}^{2}{\beta}^{2}cd+40\,{p}^{
4}b+2\,{p}^{2}a+12\,{k}^{2}{p}^{2}b )=0,\\[0.3cm]

\lambda\, (2\,{p}^{2}a+{k}^{2}a-{\beta}^{4}cd-{\beta}^{2}c+{k}^{
4}b+16\,{p}^{4}b-\omega+12\,{k}^{2}{p}^{2}b )=0,\\[0.3cm]

\beta\,b{k}^{4}-\beta\,\omega-{\beta}^{3}c-2\,ap\lambda\,k-{\beta}^{5
}cd-s\lambda-4\,bp\lambda\,{k}^{3}-8\,b{p}^{3}k\lambda+\beta\,a{k}^{2}=0.

\end{array}
\]
Solving it we obtain one set of nontrivial solution,
\begin{equation}\label{hnls_li06}
\begin{array}{l}

  s\,=\,8 b p k({k}^{2}+{p}^{2}),\,\,\,\,

  \beta\,=\,-{\dfrac {k\lambda}{p}},\,\,\,\,

  \omega=2\,{p}^{2}a+3\,{k}^{2}a+37\,{k}^{4}b+52\,{k}^{2}{p}^{2}b+16\,{p
}^{4}b,\\[0.4cm]

  c=-{\dfrac {2{p}^{2} (30\,b{k}^{2}+20\,b{p}^{2}+a )}{{
\lambda}^{2}}},\,\,\,\,

  d=-\,{\dfrac {12b{p}^{2}}{{\lambda}^{2} (30\,b{k}^{2}+20\,b{p}^{2}
+a )}} .
\end{array}
\end{equation}

From (\ref{hnls_li01}), (\ref{hnls_li03}) and (\ref{hnls_li06}), we
obtain the optical solitary wave of Eq.(\ref{hnls_01}),
\[
\begin{array}{l}
   q_1(x,t)\,=\,\left(-{\dfrac {i\,k\lambda}{p}}\,+\,\lambda\,\tanh(p\,x+8bpk({k}^{2}+{p}^{2})\,t)\right)\, {\rm
e}^{i\,\left(k\,x-(2\,{p}^{2}a+3\,{k}^{2}a+37\,{k}^{4}b+52\,{k}^{2}{p}^{2}b+16\,{p
}^{4}b)\,t+\theta\right)},
\end{array}
\]
where $p, k$ are determined by the last two identities of
Eq.(\ref{hnls_li06}). From (\ref{hnls_li03}) and (\ref{hnls_li06}),
the amplitude of the complex envelope function $E(x,t)$ reads,
\[
\begin{array}{l}
  |E(x,t)|\,=\,\left\{\dfrac{k^2\lambda^2}{p^2}\,+\,
\lambda^2\,\tanh^2(p\,x+8bpk({k}^{2}+{p}^{2})\,t) \right\}^{1/2},
\end{array}
\]
which may approach nonzero when the time variable approaches
infinity.

\section*{3.\,\,A series of exact solutions for Eq.(\ref{hnls_01}) by using F-expansion
method}

\mbox{}\hskip 0.6cm We suppose that the solution of (\ref{hnls_01})
is of the form
\begin{equation}\label{hnls_be09}
q(x,t)\,=\,P(\tau)\,{\rm
e}^{i\,\eta},\,\,\,\tau=B(x-\nu\,t),\,\,\,\eta=(-\kappa\,x+\omega\,t+\theta),
\end{equation}
where $P(\tau)$ is a real function, and $B, \nu, \kappa, \omega$ are
real constants to be determined. Substituting Eq.(\ref{hnls_be09})
to Eq.(\ref{hnls_01}) and separating the real and imaginary parts,
one may obtain the following equations,
\begin{equation}\label{hnls_09}
 -\,B(\nu+2a\kappa+4b\kappa^3)\,P^{\prime}\,+\,4b\,\kappa\,B^3\,P^{\prime\prime\prime}\,=\,0,
\end{equation}
\begin{equation}\label{hnls_10}
(\omega+a\,\kappa^2+b\kappa^4)\,P\,-\,cP^3\,-\,c\,d\,P^5\,-\,B^2(a+6b\kappa^2)\,P^{\prime\prime}
\,+\,bB^4\,P^{\prime\prime\prime\prime}\,=\,0.
\end{equation}
The linear ordinary differential equation (\ref{hnls_09}) has no
solitary wave solutions, thus we have to take $\kappa=\nu=0$. In
this case Eq.(\ref{hnls_09}) is satisfied identically, and
Eq.(\ref{hnls_10}) becomes,
\begin{equation}\label{hnls_11}
\omega\,P\,-\,cP^3\,-\,c\,d\,P^5\,-\,a\,B^2\,P^{\prime\prime}
\,+\,bB^4\,P^{\prime\prime\prime\prime}\,=\,0.
\end{equation}

Eq.(\ref{hnls_11}) can be solved by using the F-expansion
method\cite{zhouyb_pla_2003}-\cite{yomba_pla_2008}. According to the
F-expansion method, we suppose,
\begin{equation}\label{hnls_13}
  P(\tau)\,=\,\sum\limits_{i=0}^{n}\,A_i\,F^i(\tau),\,\,A_n\,\neq 0,
\end{equation}
where $A_i(i=0,\cdots, n)$ are real constants to be determined, the
integer $n$ is determined by balancing the linear highest order term
and nonlinear term. And $F(\tau)$ in (\ref{hnls_13}) satisfies,
\begin{equation}\label{hnls_14}
  \left(\dfrac{{\rm d}\,F(\tau)}{{\rm d}\,\tau}\right)^2\,=\,h_0\,+\,h_2\,F(\tau)^2\,+\,h_4\,F(\tau)^4,
\end{equation}
where $h_0, h_2, h_4$ are real constants. By balancing the linear
highest order derivative term $P^{\prime\prime\prime\prime}$ with
nonlinear term $P^5$ in Eq.(\ref{hnls_11}), we find $n=1$. Thus
Eq.(\ref{hnls_13}) becomes,
\begin{equation}\label{hnls_15}
  P(\tau)\,=\,A_0\,+\,A_1\,F(\tau).
\end{equation}

Substituting Eq.(\ref{hnls_15}) into Eqs.(\ref{hnls_11}) along with
Eq.(\ref{hnls_14}), collecting all terms with the same power of
$F^j(\tau)(j=0,\cdots,5)$, and equating the coefficients of these
terms yields a set of algebraic equations with respect to $A_0$,
$A_1$, $B$, $\omega$, $a$, $b$, $c$, $d$, $h_0$, $h_2$, $h_4$:
\[
\begin{array}{l}

\,dc{A_{{1}}}^{4}A_{{0}}\,=\,0,\hskip 0.3cm
\omega\,A_{{0}}-c{A_{{0}}}^{3}-dc{A_{{0}}}^{5}\,=\,0,\\[0.3cm]

10\,dc{A_{{1}}}^{2}{A_{{0}}}^{3}+3\,c{A_{{1}}}^{2}A_{{0}}\,=\,0,\hskip
0.3cm
24\,bA_{{1}}{B}^{4}{h_{{4}}}^{2}\,-\,d\,c{A_{{1}}}^{5}\,=\,0,\\[0.3cm]

20\,bA_{{1}}{B}^{4}h_{{2}}h_{{4}}-10\,dc{A_{{1}}}^{3}{A_{{0}}}^{2}-c{A_{{1}}}^{3}-2\,A_{{1}}{B}^{2}ah_{
{4}}\,=\,0,\\[0.3cm]

\omega\,A_{{1}}-3\,cA_{{1}}{A_{{0}}}^{2}-A_{{1}}{B}^{2}ah_{{2}}+12\,bA_{{1}}{B}^{4}h_
{{4}}h_{{0}}+bA_{{1}}{B}^{4}{h_{{2}}}^{2}-5\,dcA_{{1}}
{A_{{0}}}^{4}\,=\,0.

\end{array}
\]
Solving the above algebraic equations, we have a set of nontrivial
solution,
\begin{equation}\label{hnls_16}
\begin{array}{l}

   A_0\,=\,0,\hskip 0.5cm
   A_1\,=\,\pm\sqrt{\dfrac{12bB^2h_4}{d(10bB^2h_2-a)}},\\[0.45cm]

  \omega={B}^{2}\,(ah_{{2}}-12\,b{B}^{2}h_{{4}}h_{{0}}-b{B}^{2}{h_{{2}}}^{2
}),\hskip 0.3cm

  c\,=\,\dfrac {d\,(10\,b{B}^{2}h_{{2}}-a)^2}{6b}.
\end{array}
\end{equation}

Special analytical solutions to Eq.(\ref{hnls_14}) exists for
certain choices of the constants $h_0$, $h_2$ and $h_4$. When
$h_0=1$, $h_2=-(1+m^2)$, $h_4=m^2$, Eq.(\ref{hnls_14}) has the
solution $F(\tau)=\sn(\tau,m)$. From Eq.(\ref{hnls_be09}) and Eq.
(\ref{hnls_15}), Eq.(\ref{hnls_01}) has the Jacobian elliptic sine
function solution,
$$
  q_2(x,t)=\pm \sqrt{-\,{\dfrac{12{B}^{2}b{m}^{2}}{d(10{B}^{2}b+10{B}^{2}b{m}^{2}
+a)}} }\,\sn(B\,x, m)\,{\rm
  e}^{i\,(-{B}^{2}({B}^{2}b+{B}^{2}b{m}^{4}+14{B}^{2}b{m}^{2}+a{m}^{2}+a
 )\,t+\theta)},
$$
where $B$ is determined by $d(10bB^2+10m^2bB^2+a)^2-6bc=0$.

When $h_0=1-m^2$, $h_2=2m^2-1$, $h_4=-m^2$, Eq.(\ref{hnls_14}) has
the solution $F(\tau)=\cn(\tau,m)$. Inserting it into
(\ref{hnls_15}) and using the transformation (\ref{hnls_be09}),
Eq.(\ref{hnls_01}) has the Jacobian elliptic cosine function
solution,
$$
  q_3(x,t)\,=\pm\,\sqrt{{\dfrac {12{B}^{2}b{m}^{2}}{d(10\,{B}^{2}b-20\,{B}^{2}b{m}^{2}
+a)}}} \,\cn(B\,x, m)\,{\rm
  e}^{i\,({B}^{2}(16{B}^{2}b{m}^{2}-16{B}^{2}b{m}^{4}-{B}^{2}b-a+2a
{m}^{2})\,t+\theta)},
$$
where $B$ is determined by $d(a+10bB^2-20bB^2m^2)^2-6bc=0$.

Some solitary wave solutions can be obtained if the modulus $m$
approaches to 1. For example, when $m \rightarrow 1$, the solution
$q_2(x,t)$ degenerates to the kink type envelope wave solution,
$$
  q_4(x,t)\,=\pm\,\sqrt{-{\dfrac {12b\,{B}^{2}}{d(20\,{B}^{2}b+a)}}}\,\tanh(B\,x)\,{\rm
  e}^{i\,(-2\,{B}^{2}(8\,{B}^{2}b+a)\,t+\theta)},
$$
where $B$ is determined by $d(a+20bB^2)^2-6bc=0$.

When $m \rightarrow 1$, the solution $q_3(x,t)$ degenerates to the
bell type envelope wave solution,
$$
  q_5(x,t)\,=\pm\,\sqrt{{\dfrac {12{B}^{2}b}{d(a-10\,{B}^{2}b)}}}\,\sech(B\,x)\,{\rm
  e}^{i\,({B}^{2}(a-{B}^{2}b)\,t+\theta)},
$$
where $B$ is determined by $d(a-10bB^2)^2-6bc=0$.

As pointed out in Ref.\cite{zhouyb_pla_2003}, Eq.(\ref{hnls_14}) has
many other Jacobi elliptic function solutions in terms of
$\dn(\xi)$, $\ns(\xi)$, $\nd(\xi)$, $\nc(\xi)$, ${\rm sc}(\xi)$,
$\cs(\xi)$, $\sd(\xi)$, $\ds(\xi)$, $\cd(\xi)$, $\dc(\xi)$ as well
as the corresponding solitary wave and trigonometric function
solutions. For simplicity, such types of solutions to
Eq.(\ref{hnls_01}) are not listed here.

\vskip 0.2cm With the aid of Maple, we have checked the solutions
$q_j(x,t)(j=1,\cdots,5)$ by putting them back into
Eq.(\ref{hnls_01}).

\section*{Acknowledgment}

\mbox{}\hskip 0.6cm This work was supported by the National Natural
Science Foundation of China ( No. 10801037).

\end{document}